\documentclass{emulateapj}
\usepackage{verbatim}
\usepackage{natbib}

\def\del#1{{}}

\usepackage{graphicx}

\newcommand{\dd}{\mathrm{d}}
\newcommand{\bra}{\langle}
\newcommand{\ket}{\rangle}
\newcommand{\ltsima}{$\; \buildrel < \over \sim \;$}
\newcommand{\lsim}{\lower.5ex\hbox{\ltsima}}
\newcommand{\gtsima}{$\; \buildrel > \over \sim \;$}
\newcommand{\gsim}{\lower.5ex\hbox{\gtsima}}

\newcommand{\CR}{{\rm CR}}

\newcommand{\rmn}{\mathrm}
\newcommand{\dps}{\displaystyle}

\newcommand{\eps}{\varepsilon}

\shorttitle{MAGIC Observation of the Perseus Galaxy Cluster}
\shortauthors{Aleksi\'c, J. et al.}

\begin{document}
\title{MAGIC gamma-ray telescope observation of the Perseus cluster of galaxies:
implications for cosmic rays, dark matter, and NGC~1275}

%
\author{
J.~Aleksi\'c\altaffilmark{a},
L.~A.~Antonelli\altaffilmark{b},
P.~Antoranz\altaffilmark{c},
M.~Backes\altaffilmark{d},
C.~Baixeras\altaffilmark{e},
S.~Balestra\altaffilmark{c},
J.~A.~Barrio\altaffilmark{c},
D.~Bastieri\altaffilmark{f},
J.~Becerra Gonz\'alez\altaffilmark{g},
W.~Bednarek\altaffilmark{h}
% ...
\\
\vspace{0.1cm}
({The MAGIC Collaboration})\\
\vspace{0.2cm}
C.~Pfrommer\altaffilmark{z,\dag}, 
A.~Pinzke\altaffilmark{1},
T.~A.~En{\ss}lin\altaffilmark{2},
S.~Inoue\altaffilmark{3} and
G.~Ghisellini\altaffilmark{4}
}
\altaffiltext{a} {IFAE, Edifici Cn., Campus UAB, E-08193 Bellaterra, Spain}
\altaffiltext{b} {INAF National Institute for Astrophysics, I-00136 Rome, Italy}
\altaffiltext{c} {Universidad Complutense, E-28040 Madrid, Spain}
\altaffiltext{d} {Technische Universit\"at Dortmund, D-44221 Dortmund, Germany}
\altaffiltext{e} {Universitat Aut\`onoma de Barcelona, E-08193 Bellaterra, Spain}
\altaffiltext{f} {Universit\`a di Padova and INFN, I-35131 Padova, Italy}
\altaffiltext{g} {Inst. de Astrof\'{\i}sica de Canarias, E-38200 La Laguna, Tenerife, Spain}
\altaffiltext{h} {University of \L\'od\'z, PL-90236 Lodz, Poland}
% ...
\altaffiltext{z} {CITA, University of Toronto, M5S 3H8 Toronto, Canada}
\altaffiltext{1} {Stockholm University, SE - 106 91 Stockholm, Sweden}
\altaffiltext{2} {Max-Planck-Institut f\"ur Astrophysik, D-85740 Garching, Germany}
\altaffiltext{3} {Kyoto University, 606-8502 Kyoto, Japan}
\altaffiltext{4} {INAF National Institute for Astrophysics, I-23807 Merate, Italy}
% ...
\altaffiltext{\dag} {Send offprint requests to F.~Zandanel (fabio@iaa.es) \& C.~Pfrommer (pfrommer@cita.utoronto.ca)}


\begin{abstract}
  The Perseus galaxy cluster was observed by the MAGIC Cherenkov telescope for a
  total effective time of 24.4~hr during 2008 November and December. The
  resulting upper limits on the $\gamma$-ray emission above 100~GeV are in the
  range of $4.6$ to $7.5\times 10^{-12}\rmn{ cm}^{-2}\,\rmn{s}^{-1}$ for
  spectral indices from $-1.5$ to $-2.5$, thereby constraining the emission
  produced by cosmic rays, dark matter annihilations, and the central radio
  galaxy NGC~1275. Results are compatible with cosmological cluster simulations
  for the cosmic-ray-induced $\gamma$-ray emission, constraining the average
  cosmic ray-to-thermal pressure to $<4\%$ for the cluster core region ($<8\%$
  for the entire cluster). Using simplified assumptions adopted in earlier work
  (a power-law spectrum with an index of $-2.1$, constant cosmic ray-to-thermal
  pressure for the peripheral cluster regions while accounting for the adiabatic
  contraction during the cooling flow formation), we would limit the ratio
  of cosmic ray-to-thermal energy to $E_\CR/E_\rmn{th}<3\%$. Improving the
  sensitivity of this observation by a factor of about 7 will enable us to
  scrutinize the hadronic model for the Perseus radio mini-halo: a non-detection
  of $\gamma$-ray emission at this level implies cosmic ray fluxes that are too
  small to produce enough electrons through hadronic interactions with the
  ambient gas protons to explain the observed synchrotron emission. The upper
  limit also translates into a level of $\gamma$-ray emission from possible
  annihilations of the cluster dark matter (the dominant mass component) that is
  consistent with boost factors of $\sim10^4$ for the typically expected dark
  matter annihilation-induced emission.  Finally, the upper limits obtained for
  the $\gamma$-ray emission of the central radio galaxy NGC~1275 are consistent
  with the recent detection by the \emph{Fermi}-LAT satellite. Due to the extremely
  large Doppler factors required for the jet, a one-zone synchrotron
  self-Compton model is implausible in this case. We reproduce the observed
  spectral energy density by using the structured jet (spine-layer) model which
  has previously been adopted to explain the high-energy emission of radio
  galaxies.
\end{abstract}


 
\section{Introduction}
Clusters of galaxies provide us with the opportunity to study an ``ecosystem'',
a volume that is a high-density microcosm of the rest of the Universe.  Clusters
of galaxies are the largest and most massive gravitationally bound systems in
the Universe, with radii of few Mpc and total masses $M \sim (10^{14}-10^{15})
M_{\odot}$, of which galaxies, gas, and dark matter (DM) contribute roughly for 5\%, 15\% and
80\%, respectively (see, e.g., \citealp{1988xrec.book.....S}; \citealp{2003ApJ...585..161K}; 
\citealp{2005RvMP...77..207V} for a general overview).
While no cluster has been firmly detected as a $\gamma$-ray source so far
\citep{2003ApJ...588..155R, 2006ApJ...644..148P, 2008AIPC.1085..569P, 
2009A&A...495...27A, 2009arXiv0907.0727T, 2009arXiv0907.3001D, 
2009arXiv0907.5000G, cangaroo_clusters,2009arXiv0911.0740V}, they are expected to be significant
$\gamma$-ray emitters on the following general grounds. (1) Clusters are
actively evolving objects and being assembled today, in the latest and most
energetic phase of hierarchical structure formation. (2) Clusters serve as
cosmic energy reservoirs for powerful sources such as radio galaxies and
supernova-driven galactic winds. (3) Finally, clusters contain large amounts
of gas with embedded magnetic fields, often showing direct evidence for shocks
and turbulence as well as relativistic particles. For recent reviews regarding non-thermal 
processes in clusters as well as numerical simulations, see \cite{2007IJMPA..22..681B}
and \cite{2008SSRv..134..311D}.

In the cosmological hierarchic clustering model, large-scale structures grow
hierarchically through merging and accretion of smaller systems into larger
ones, and clusters are the latest and most massive objects to form 
(e.g.,~\citealp{1993ppc..book.....P}). 
Recently, high resolution X-ray observations by \emph{Chandra} and \emph{XMM-Newton}
orbiting telescopes provided confirmation of this picture 
(e.g.,~\citealp{2002ARA&A..40..539R,2005RvMP...77..207V}).  
During the course of cluster assembly, energies of the order of the final gas binding 
energy $E_{b} \sim 3 \times (10^{61}-10^{63}) $~erg should be dissipated through merger 
and accretion shocks (collectively called ``structure formation shocks'') as well as
turbulence. The energy is expected to be dissipated on a dynamical timescale of
$\tau_\rmn{dyn}\sim 1$~Gyr. Hence the corresponding rates of energy release are
$L \sim (10^{45} - 10^{47}) $~erg~s$^{-1}$, so even a small fraction of this energy
channeled into non-thermal particles can be of major observable consequence.
Shocks and turbulence are also likely to accelerate non-thermal electrons and
protons to high energies (e.g.,~\citealp{1977ApJ...212....1J,
1987A&A...182...21S,2001MNRAS.320..365B,2001ApJ...559...59M,2001ApJ...562..233M,2002ApJ...577..658O,
2002MNRAS.337..199M,2003MNRAS.342.1009M,2002mpgc.book....1S,2004MNRAS.350.1174B,2005ApJ...628L...9I,
2007ApJ...670L...5B,2007MNRAS.378..245B,2007MNRAS...378..385P,2008MNRAS.385.1211P,2008MNRAS.385.1242P,
2009arXiv0912.0545F}).

Clusters are also home to different types of
energetic outflows, and the intra-cluster medium (ICM) can function as an
efficient energy reservoir. Most clusters are seen to harbor radio galaxies
around their central regions, whose large, powerful jets of relativistic plasma
are interacting vigorously with the ICM \citep{1998ApJ...501..126H,
2003astro.ph..1476F, 2006MNRAS.366..417F}.  A crude estimate of the total energy output by a single
powerful radio galaxy is $E_{RG} \sim (10^{60} - 10^{62}) $~erg, taking reasonable
values for the kinetic luminosity $L_{RG} \sim (10^{45}-10^{46}) $ erg~s$^{-1}$ and
effective duration of activity $t_{RG} \sim (10^{7}-10^{8})$ yr \citep{2007ARA&A..45..117M}. 
The integrated output from the whole cluster radio galaxy population should be even greater
\citep{1997ApJ...477..560E,1998A&A...333L..47E,2001ApJ...562..618I}.  Although
rarely seen in present-day clusters, another source which should have been
active in the past are galactic winds, i.e.~outflows driven by the joint action
of numerous supernovae \citep{1996SSRv...75..279V}. Taking the observed mass of
Fe in the ICM to be $M_{Fe,ICM} \sim 3 \times (10^{9}-10^{10}) M_\odot$, the
energy and Fe mass ejected by each supernovae to be respectively $E_{SN} \sim
10^{51}$ erg and $M_{Fe,SN} \sim 0.1 M_\odot$, and an outflow efficiency
$\xi_{GW} \sim 0.1$ \citep{2005ARA&A..43..769V}, we estimate the total galactic wind 
energy output to be $E_{GW} \sim \xi_{GW} E_{SN}/M_{Fe,SN} M_{Fe,ICM} \sim 3 \times
(10^{60}-10^{61})$ erg.  In any case, along with dumping energy, these sources
can inject substantial quantities of non-thermal particles into the ICM, or
could have done so in the past.

Faraday rotation measurements provide a powerful tool to probe the strength of
the intra-cluster magnetic fields \citep{1991ApJ...379...80K} and even their
distribution \citep{2001ApJ...547L.111C}, resulting in the ICM now being known
to be permeated by magnetic fields with strengths $B \sim (1-10)$~$\mu$G
\citep{2002ARA&A..40..319C,2005A&A...434...67V}, which allow for particle
acceleration in shocks up to $\gamma$-ray emitting energies.  Observations of
radio halos and radio relics have already established that synchrotron emitting
electrons with energies reaching $\sim10$~GeV are present in at least some
clusters \citep{2003ASPC..301..143F,2008SSRv..134...93F}, although their precise
origin is still unclear. Similar populations of electrons but with harder
spectra may produce $\gamma$-rays efficiently via inverse Compton (IC)
up-scattering of the cosmic microwave background \citep{2000Natur.405..156L,
  2000ApJ...545..572T,
  2002MNRAS.337..199M,2003MNRAS.342.1009M,2008SSRv..134..191P}.  Observations in
the hard X-ray regime may suggest the presence of a non-thermal component due to
the IC scattering of cosmic microwave photons by relativistic electrons (see
\citealp{2008SSRv..134...71R} for a recent review). However,
\cite{2009ApJ...690..367A} found no evidence of a hard tail above the thermal
emission in a \emph{Swift}/Burst Alert Telescope (BAT) sample of clusters. The ICM gas should also provide
ample target matter for inelastic collisions leading to pion-decay $\gamma$-rays
\citep{1996SSRv...75..279V,1997ApJ...477..560E, 2003A&A...407L..73P,
  2004A&A...413...17P,2008MNRAS.385.1211P,2008MNRAS.385.1242P} as well as
secondary electron injection \citep{1980ApJ...239L..93D,1982AJ.....87.1266V,
  1999APh....12..169B,2000A&A...362..151D,2004A&A...413...17P,2007ApJ...663L..61F,2008MNRAS.385.1242P}.
The magnetic fields play another crucial role by confining non-thermal protons
within the cluster volume for longer than a Hubble time, i.e.~any protons
injected into the ICM accumulates throughout the cluster history
\citep{1996SSRv...75..279V,1997ApJ...487..529B}.

Galaxy clusters present very large M/L ratios and considerable overdensities, which are crucial for indirect
DM searches.  Despite the fact that they are not as near as other potential DM candidates, 
as the dwarf spheroidal galaxies \citep{2008ApJ...679..428A,2009ApJ...697.1299A}, the large 
DM masses of clusters could make them ideal laboratories also for the search of a DM annihilation $\gamma$-ray
signal (\citealp{2008arXiv0812.0597J}; \citealp{2009arXiv0905.1948P}).

In this paper we report the results of the Perseus cluster observation performed by the Major Atmospheric Gamma 
Imaging Cherenkov (MAGIC) telescope for a total effective time of $24.4$~hr during 2008 November and December. 
In Section~2, we explain the physical motivations why we chose Perseus over other galaxy
clusters and present its main characteristics. In Section~3, we briefly introduce the MAGIC telescope. 
We then describe the Perseus data sample, the analysis, and the obtained flux upper limits. We 
discuss the implications for the cosmic ray (CR) pressure and the possible DM annihilation-induced
$\gamma$-ray emission in Section~4 and 5, respectively. In Section~6, we discuss the implications
for the jet emission model of the central radio galaxy NGC~1275. Finally, in Section~7, we summarize 
our conclusions. All cluster masses and luminosities are scaled to the currently favored
value of Hubble's constant $H_{0} = 70$~km~s$^{-1}$~Mpc$^{-1}$.

\section{Target Selection and Preliminaries}
\label{sec:target}
The Perseus cluster, also called A426, is at a distance of 77.7~Mpc ($z =
0.018$). It is the brightest X-ray cluster \citep{1992MNRAS.258..177E} and hosts
a massive cooling flow with high central gas densities of $0.05\mbox{ cm}^{-3}$
(see Table~\ref{tab:data}).  Perseus furthermore hosts a luminous radio
mini-halo -- diffuse synchrotron emission that fills a large fraction of the
cluster core region -- and shows a source extension of $\sim 200$~kpc
\citep{1990MNRAS.246..477P}. This radio mini-halo is well modeled by the
hadronic scenario where the radio emitting electrons are produced in hadronic
CR proton interactions with ambient gas protons requiring only a
very modest fraction of a few percent CR pressure relative to thermal pressure
\citep{2004A&A...413...17P}. In particular, the similarity of the thermal X-ray
emission to that of the radio mini-halo comes about naturally as both processes
scale with the number density squared.  An alternative model for the radio
emission has been proposed by \citet{2002A&A...386..456G} which explains the
radio mini-halo by re-acceleration of relativistic electrons through second-order 
interactions with magneto-hydrodynamic (MHD) turbulence.  However, it
remains to be shown whether the necessary turbulent energy density can be
provided throughout the entire cooling flow region of Perseus.  These conditions
provide high target densities for hadronic CRp-p interactions and enhance the
resulting $\gamma$-ray flux.

The Perseus galaxy cluster was carefully chosen over other nearby clusters after
considering the expected $\gamma$-ray emission from the pion-decay and DM
annihilation. Moreover, the central radio galaxy NGC~1275 is expected to be a
promising GeV-TeV target, and hence is another strong motivation to observe
this cluster. In the following subsections, we detail our considerations.

\begin{table*}[hbt!]
\begin{center}
\caption{\label{tab:data}Properties of the Perseus Galaxy Cluster}
\begin{tabular}{cccccccc}
\hline\hline
\phantom{\Big|} 
  $z$ & $D_\rmn{lum}$ [Mpc] & $R_{200}$ [Mpc] & $M_{200}$ [M$_{\odot}$]& 
  $L_{X,0.1-2.4}$ [erg~s$^{-1}$] & $T_X$ [keV] & $L_{\nu=1.4}$ [erg~s$^{-1}$~Hz$^{-1}$]\\
\hline 
\phantom{\Big|}
0.0183 & 77.7 & 1.9 & $7.71 \times 10^{14}$ & $8.31 \times 10^{44}$ & 6.8 & $3.38 \times 10^{31}$ \\
\hline\hline
\end{tabular}
\end{center}
{\bf Notes.} Data taken from \cite{2002ApJ...567..716R}, \cite{1990MNRAS.246..477P}, and \cite{2003ApJ...590..225C}.
\end{table*}

\subsection{Cosmic-Ray-Induced Emission}
\label{sec:target_cluster}
In the course of this work, we used cosmological simulations of the formation of
galaxy clusters to inform us about the expected spatial and spectral
characteristics of the CR induced $\gamma$-ray emission. A clear
detection of the IC emission from shock-accelerated CR electrons will be
challenging for Imaging Atmospheric Cherenkov Telescopes (IACTs) due to the
large angular extent of these accretion shocks that subtend solid angles
corresponding to up to six virial radii. For these instruments, the spatially
concentrated pion-decay $\gamma$-ray emission resulting from hadronic CR
interactions that dominates the total $\gamma$-ray luminosity
\citep{2008MNRAS.385.1211P,2008MNRAS.385.1242P} should be more readily
detectable than the emission from the outer region.

To address the question of universality and predictability of the expected
$\gamma$-ray emission, we simulated a sample of 14 galaxy clusters that span 
one and a half decades in mass and show a variety of dynamical states ranging
from relaxed cool core clusters to violent merging clusters (details are given
in Sect.~\ref{sec:simulations}). In order to find the most promising target
cluster in the local Universe for detecting the pion-decay emission, we computed
the scaling relations between the $\gamma$-ray luminosity and cluster mass of our
sample \citep{2008MNRAS.385.1242P} and used these to normalize the CR-induced
emission of all clusters in a complete sample of the X-ray brightest clusters
(the extended HIFLUGCS catalog; \citealp{2002ApJ...567..716R}).  This favors
a high-mass, nearby galaxy cluster with a scaling $M_{200}^\beta / D_\rmn{lum}^2$,
where $M_{200}$ is the virial mass\footnote[1]{We define the virial mass
  $M_\Delta$ and the virial radius $R_\Delta$ as the mass and radius of a sphere
  enclosing a mean density that is $\Delta=200$ times the critical density of
  the Universe.}, $D_\rmn{lum}$ is the luminosity distance, and $\beta \simeq 1.32$
is a weakly model-dependent scaling parameter that provides the rank ordering
according to the brightness of each individual cluster
\citep{2008MNRAS.385.1242P}.  As a second criterion, we required low zenith
angle observations, i.e below $35^\circ$, that ensure the lowest possible energy
thresholds and the maximum sensitivity for the detector.  We carefully modeled
the most promising targets, accounting for the measured gas density and
temperatures from thermal X-ray measurements while assuming a constant
CR-to-thermal gas ratio \citep{2004A&A...413...17P}. Cluster-wide extended radio
synchrotron emission that informs about present high-energy processes was
additionally taken into account before we selected the Perseus cluster as our
most promising source.  Although other clusters showed a somewhat higher
$\gamma$-ray flux in our simulations (e.g., Ophiuchus), the facts that Perseus is
observable at low zenith angles and that the expected emission is more spatially
concentrated make it the best suited target for this observation.

\subsection{Dark Matter Content}
\label{section:target_DM}
Typically up to 80\% of the total mass of a galaxy cluster is in the form of
non-baryonic DM. Since the DM annihilation $\gamma$-ray signal is expected to 
be proportional to the integrated squared DM density along the line of sight
\citep{2004PhRvD..69l3501E,2006PhRvD..73f3510B}, it is obvious that galaxy
clusters could be good candidates to look for DM as well.  This is true despite
the fact that they are located at much larger distances than other potential DM
candidates, such as dwarf spheroidal galaxy satellites of the Milky Way or the
Galactic Center. One obvious reason is the huge amount of DM hosted by clusters 
compared with the rest of candidates. Perseus, for example, is located $\sim$1000 
times farther than Milky Way dwarfs, but it contains roughly six orders of magnitude
more DM than the Willman~1 dwarf galaxy, one of the most promising DM candidates
according to recent work \citep{2007PhRvD..75h3526S,2009ApJ...697.1299A}. Additionally,
the presence of substructures could be of crucial importance. Substructures in clusters 
may significantly enhance the DM signal over the smooth halo, while we do not expect this 
to be of special relevance for dwarf galaxies since their outer regions are severely affected 
by tidal stripping (\citealp{2009arXiv0905.1948P}; S\'anchez-Conde et al. 2010, in prep.).

Essentially, the annihilation flux is proportional to the product of two
parameters (see e.g., \citealp{2004PhRvD..69l3501E} for details): a first one
that captures all the particle physics (DM particle mass, cross section, etc.),
which we will label as $f_{\mathrm{SUSY}}$, and a second one,
$J_{\mathrm{astro}}$, that accounts for all the astrophysical considerations (DM
distribution, telescope point-spread function (PSF), etc.). The particle physics factor just acts as a
normalization in the expected annihilation flux, so we can neglect it when
performing a comparative study -- as we are doing in this section. Concerning
the astrophysical factor, the DM distribution is commonly modeled with radial
density profiles of the form $\rho(r) = \rho_s / [(r/r_s )^\gamma~(1 +
(r/r_s)^\alpha)^{(\beta-\gamma)/\alpha}]$, where $\rho_s$ and $r_s$ represent a
characteristic density and a scale radius respectively
\citep{1998ApJ...502...48K}. These density profiles are well motivated by
high-resolution N-body cosmological simulations. Here we adopt the
Navarro-Frenk-White (\citealp{1997ApJ...490..493N}, hereafter NFW) DM density
profile, with ($\alpha$,$\beta$,$\gamma$) = (1, 3, 1).  For an NFW profile, 90\%
of the DM annihilation flux comes from the region within $r_s$, so that the
corresponding integrated luminosity is proportional to $r_s ^3 \rho_s^2$.  We
can derive $r_s$ and $\rho_s$ for Perseus, assuming $M_{200}=7.7 \times 10^{14}$
$M_\odot$ (as given in Table~\ref{tab:data}) and a concentration of $\sim$6 (as
given by the \citealp{2001MNRAS.321..559B} virial mass-concentration scaling
relation). We obtain $r_s=0.384$~Mpc and $\rho_s=1.06 \times
10^{15}$~M$_\odot$~Mpc$^{-3}$, which translates into a {\it total} value of
$J_{\mathrm{astro}} \sim 1.4 \times 10^{16}$~GeV$^2$~cm$^{-5}$ for the scale
radius region.  
In the case of Coma, although
slightly ($\sim$15\%) more massive than Perseus, the fact that it is located
significantly farther (101~Mpc) translates into a slightly lower annihilation
flux. Virgo, only 17~Mpc away from us, gives a larger DM annihilation flux, but
here the large extension of the region from which most of the annihilation flux
is expected to come compared with Perseus ($r_s\sim1^{\circ}.2$ and
$r_s\sim0^{\circ}.3$, respectively) could represent an obstacle from the
observational point of view. Source extension is of special relevance
for single-telescope IACTs, for which point-like sources (sources with an
angular extension smaller than or similar to the telescope PSF) are more readily
observable.

\subsection{The NGC~1275 Radio Galaxy}
The central NGC~1275 radio galaxy is another strong motivation for $\gamma$-ray
observations of the Perseus galaxy cluster. The detection at TeV energies of the
radio galaxies M~87 \citep{2006Sci...314.1424A} and Centaurus~A
\citep{2009ApJ...695L..40A} has forced a substantial revision of the paradigm
whereby very high energy (VHE) emission is a characteristic property of highly relativistic jets
closely aligned with the line of sight, establishing radio galaxies as a new
class of VHE $\gamma$-ray emitters. Note that NGC~1275 has various characteristics 
in common with Centaurus~A which has also been interpreted as a possible source of 
ultra-high-energy CRs \citep{2009MNRAS.393.1041H}.

The NGC~1275 radio galaxy is the brightest radio source in the northern sky. Its
jet inclination angle seems to increase from $10^{\circ}-20^{\circ}$ at milliarcsecond
scales up to $40^{\circ}-60^{\circ}$ at arcsecond scales \citep{2006MNRAS.366..758D}.  Note
that NGC~1275 was classified as a blazar by \cite{1980ARA&A..18..321A} because
of its optical polarization, and it has been seen to vary in the optical on time
scales of a day \citep{1979ApJ...230L.141G}. All these elements are promising
from the point of view of the TeV detectability, since they suggest that the
emission region is located at the base of the jet.  In these conditions, in the
scenario based on the structured jet model \citep{2008MNRAS.385L..98T}, we
expected VHE emission from the layer of the jet at a level detectable by MAGIC.

\section{MAGIC Observation and Results}
The MAGIC telescope is
located on the Canary Island of La Palma (2200 m asl, $28.45^\circ$N,
$17.54^\circ$W). With a primary mirror diameter of 17~m, it is currently 
the largest IACT. CRs impinging the Earth atmosphere originate atmospheric
showers that in turn produce Cherenkov light. The ultra-violet Cherenkov 
flashes are reflected in the focal plane of the telescope, where a camera of 577 
photomultipliers records the resulting images. MAGIC reconstructs the 
incoming $\gamma$-ray directions with an accuracy of about $0^{\circ}.1$
and achieves an energy resolution above 150~GeV of about $20$\% 
(see \citealp{2008ApJ...674.1037A}; \citealp{2009APh....30..293A} for details).

\subsection{Observation and Analysis}
\label{sec:data}
MAGIC observed the Perseus cluster for $33.4$~hr during 2008 November 
and December, at zenith angles between $12^\circ$ and $32^\circ$, which 
guarantees the lowest energy threshold. The observation was performed in the 
false-source tracking (wobble) mode \citep{1994APh.....2..137F} pointing alternatively 
to two different sky directions, each at a $0^{\circ}.4$ distance from the nominal target
position.

The main background for Cherenkov telescopes is due to the hadronic CRs
and the night sky background. Our standard analysis procedure is as follows 
\citep[for a detailed description, see][]{2008ApJ...674.1037A}: data calibration and extraction 
of the number of photoelectrons per pixel are done \citep{2008NIMPA.594..407A}. This is followed 
by an image cleaning procedure using the amplitude and timing information of the calibrated signals.
Particularly, the arrival times in pixels containing $>6$ photoelectrons (core pixels) are 
required to be within a time window of 4.5\,ns and for pixels containing $>3$ photoelectrons 
(boundary pixels) within a time window of 1.5\,ns from a neighboring core pixel. For the surviving pixels 
of each event, the shower parameters are reconstructed using the Hillas parameterization
algorithm \citep{1985ICRC....3..445H}. Hadronic background suppression is achieved using
a multivariate method called Random Forest \citep{random_forest,2008NIMPA.588..424A}, 
which uses the Hillas parameters to define an estimator called \emph{hadronness} (it runs from 0 
for gammas to 1 for hadrons) by comparison with Monte Carlo (MC) $\gamma$-ray simulations. 
Moreover, the Random Forest method is used for the energy estimation of a reconstructed shower. 
The gamma/hadron ($g/h$) separation in the analysis was optimized on a sample of well-understood 
Crab Nebula data, which is commonly accepted as a standard reference source for 
VHE astronomy.

Part of the data has been rejected mainly due to the bad weather conditions during 
some observation days. The total data rejected amount to $\sim 27$\%, resulting in 
$24.4$~hr effective observation time of very high data quality. Independent 
cross-checks were performed on the data giving compatible results.

\subsection{Results}
\label{sec:results}
\begin{figure}[hbt!]
\centering
\includegraphics[width=0.95\linewidth]{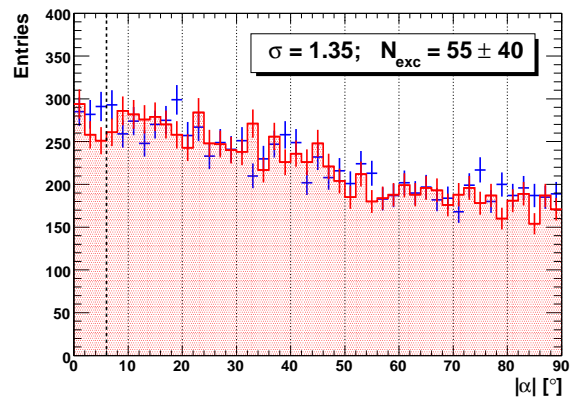}
\caption{Perseus $\alpha$-plot as seen by MAGIC in $24.4$~hr 
  above 250~GeV using a hadronness cut of $<0.05$. The blue crosses represent 
  the signal, and the red shaded region is the background. The vertical black dotted line 
  represents the fiducial region $\alpha<6^\circ$ where the signal is expected. 
  Only events above $250$~GeV are displayed since the best integral sensitivity,   
  around $1.6\%$ of Crab, is obtained from a Crab Nebula data sample in this energy range.
  \label{fig:alpha}}  
\end{figure}

Given the good data quality and the low zenith angles of observation, the analysis energy threshold results to 
be 80~GeV. Beyond this threshold, no significant excess of $\gamma$-rays above the background was detected 
in 24.4~hr of observation.
In Figure \ref{fig:alpha}, the $\alpha$-plot for energies above $250$~GeV, where the best integral sensitivity
is obtained from a Crab Nebula data sample, is reported. 
The $\alpha$-parameter is defined as the angular distance between the shower image 
main axis and the line connecting the observed source position in the camera and the image barycenter. 
Background events are isotropic in nature, and thus produce randomly oriented shower images. 
This results in a more or less smooth event distribution in the $\alpha$-plot. 
The $\gamma$-ray events due to the source, on the other hand, are predominantly aligned to the 
observed position in the camera. For a detected source, this results in a significant excess of events at small $\alpha$.  
A fiducial region $\alpha<6^\circ$ and a hadronness cut of $0.05$ are chosen by optimizing the analysis 
on a Crab Nebula data sample.

\begin{figure}[hbt!]
\centering
\includegraphics[width=0.95\linewidth]{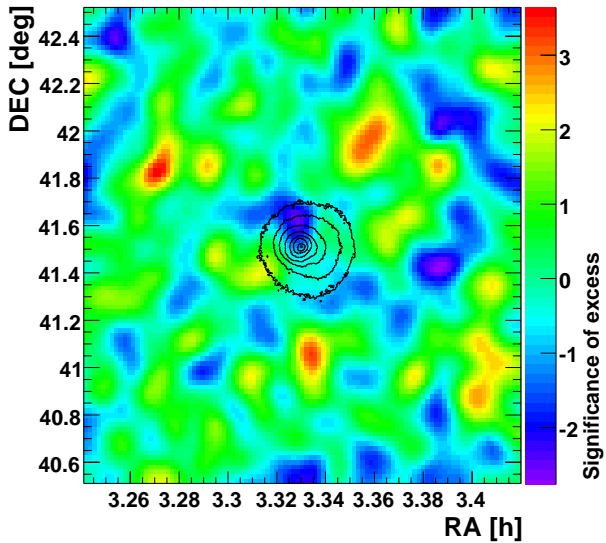}
\caption{Significance map for events above 150~GeV in the observed 
    Perseus cluster sky region. The significance distribution is consistent with background
    fluctuations. Black contours from \emph{XMM-Newton} observations in the X-ray band
    \citep{2003ApJ...590..225C} are also shown. The angular extent of the outermost 
    contours is approximately $0^{\circ}.45$, which corresponds to $\sim 610$~kpc.
    \label{fig:skymap}}  
\end{figure}

In Figure \ref{fig:skymap}, the significance map for events above 150~GeV in the observed 
sky region is shown. The source-independent DISP method has been used. This implies 
the rise of the energy threshold from 80~GeV to around 150~GeV (see \citealp{2005ICRC....5..363D} 
for a detailed description). The significance distribution in the map is consistent with background
fluctuations. In Figure \ref{fig:skymap}, X-ray contours from the \emph{XMM-Newton} observations  
\citep{2003ApJ...590..225C} are also shown.

The significance was calculated according to Eqn.~17 of
\cite{1983ApJ...272..317L} and upper limit estimation is performed using the
Rolke method \citep{2005NIMPA.551..493R}.  The upper limits in number of excess
events are calculated with a confidence level of $95$\%.  For the upper limit
calculation, a systematic uncertainty of $30$\% in the energy estimation and
effective area calculation is taken into account.  Our systematic error budget
is obtained by adding up the individual contributions in quadrature. The
different sources of systematic uncertainties are mainly related to the
differences between the real experimental conditions and the simulated ones (see
\citealp{2008ApJ...674.1037A} for a detailed discussion on the systematic errors).
The photon flux upper limit is finally reconstructed for a general $\gamma$-ray
spectrum as described in \cite{2009ApJ...697.1299A}.

In sections~\ref{subsec:CR} and \ref{subsec:DM}, we will discuss the implications
of this observation for the CR and DM annihilation-induced $\gamma$-ray flux,
respectively. Using the true density profile as obtained by X-ray measurements
\citep{2003ApJ...590..225C}, we will be able to model the spatial
characteristics of the CR-induced $\gamma$-ray signal. Our simulations indicate
that 60\% of the total $\gamma$-ray flux are contained within a circle of radius
$r_{0.6} = 0^{\circ}.15$ (this angular scale corresponds to a physical radius of
200~kpc). We then compare the flux from within this region to the upper limits.
As the characteristics of the considered emission region are close to a point
source we use point-like upper limits. The same conclusion is valid also for the
DM annihilation signal. In this case, as explained in
section~\ref{section:target_DM}, the $90$\% of the expected emission is coming
form the scale radius region. For Perseus, we obtained $r_{s}\sim0^{\circ}.3$ which
is somewhat extended compared to the telescope angular resolution. However, the
fact that the NFW profile is very steep implies that the main DM emission comes
from the core of the source that can be considered approximately point-like
compared to our angular resolution.

To compute flux upper limits, we assume specific spectral indices that have been
motivated by an astrophysical scenario in mind (see the following
sections). This ``scenario-guided'' approach allows us to provide the tightest
limits on physically motivated parameters and underlying astrophysical
models.  In the next sections we will consider flux upper limits computed using
a power-law $\gamma$-ray spectrum with spectral indices $\Gamma$ of $-1.5$,
$-2.2$, and $-2.5$. In Table~\ref{tab:UL_resume}, the corresponding integral flux
upper limits for energies above 100~GeV are listed.

\begin{table}[hbt!]
\begin{center}
\caption{\label{tab:UL_resume} Integral Flux Upper Limits Above 100~GeV}
\begin{tabular}{cc}
\hline\hline
\phantom{\Big|}
$\Gamma$ & F$_{UL}$ [$\times10^{-12}$~cm$^{-2}$~s$^{-1}$]\\
\hline
-1.5  &  4.63 \\
-2.2  &  6.55 \\
-2.5  &  7.52 \\
\hline\hline
\end{tabular}
\end{center}
{\bf Notes.} Integral flux upper limits are listed for a power-law $\gamma$-ray spectrum with spectral 
index $\Gamma$ for energies above 100~GeV. The corresponding upper limit
for the number of excess events is $186$.
\end{table} 

In Section~\ref{subsec:CR}, we will use an integral flux upper limit set above given
energy thresholds in order to trace the energy range where we can better
constrain the models. In Table~\ref{tab:int_UL}, the obtained integral flux upper
limits for $\Gamma=-2.2$ are shown. Note that we do not compute integral upper
limits above 80~GeV (as we have not shown a cumulative $\alpha$-plot for
energies above this value). This is because the $g/h$ separation for events
below 100~GeV works in a substantially different way with respect to the higher
energy events. Therefore, we analyze separately the events below 100~GeV and the
events of higher energy, with different sets of analysis cuts.

Finally, for completeness, in Table~\ref{tab:dif_UL} the differential flux upper
limits for the assumed spectral indices are shown in different energy
intervals. Spectral energy density (SED) upper limits can also be obtained from
those differential flux upper limits, as done in Section~\ref{subsec:AGN} 
discussing the observation implications for the radio galaxy NGC~1275.

\begin{table}[hbt!]
\begin{center}
\caption{\label{tab:int_UL} Integral Flux Upper Limits for a Power-law $\gamma$-ray Spectrum with Spectral 
Index $\Gamma=-2.2$ Above a Given Energy Threshold $E_{\mathrm{th}}$.}
\begin{tabular}{cc}
\hline\hline
\phantom{\Big|}
E$_{\mathrm{th}} [\mathrm{GeV}]$ & F$_{UL}$ [$\times10^{-12}$~cm$^{-2}$~s$^{-1}$]\\
\hline
100  & 6.55 \\
130  & 6.21 \\
160  & 6.17 \\
200  & 5.49 \\
250  & 4.59 \\
320  & 3.36 \\
400  & 1.83 \\
500  & 1.39 \\
630  & 0.72 \\
800  & 0.65 \\
1000 & 0.47 \\   
\hline\hline
\end{tabular}
\end{center}
\end{table}

\begin{table*}[hbt!]
\begin{center}
\caption{\label{tab:dif_UL}Differential Flux Upper Limits}
\begin{tabular}{cccccccc}
\hline\hline
\phantom{\Big|}
$\Gamma$ & [80-100] & [100-160] & [160-250] & [250-400] & [400-630] & [630-1000] & [1000-10000] \\
\hline
-1.5 & $130.7$  & $23.6$ & $12.6$ & $4.33$ & $0.865$ &  $0.168$ & $0.015$\\
-2.2 & $144.8$  & $25.3$ & $13.2$ & $4.53$ & $0.897$ &  $0.174$ & $0.018$\\
-2.5 & $150.6$  & $25.8$ & $13.3$ & $4.57$ & $0.903$ &  $0.176$ & $0.018$\\
\hline\hline
\end{tabular}
\end{center}
{\bf Notes.} Differential flux upper limits are listed in units of $10^{-11}$~cm$^{-2}$~s$^{-1}$~TeV$^{-1}$ for 
a power-law $\gamma$-ray spectrum with spectral index $\Gamma$ in energy ranges in units of GeV.
\end{table*}

\subsection{Comparison to Previous Observations}
\label{subsec:UL_diss}
There are few existing IACT observations of galaxy clusters 
\citep{2006ApJ...644..148P,2008AIPC.1085..569P,2009A&A...495...27A, 
2009arXiv0907.0727T,2009arXiv0907.3001D,2009arXiv0907.5000G,
cangaroo_clusters,2009arXiv0911.0740V}. In section~\ref{subsec:comparison}, 
we will compare the limits on the CR-to-thermal pressure 
obtained by other IACTs with those derived in this work. However, 
there are two observations of the Perseus galaxy cluster made 
by WHIPPLE \citep{2006ApJ...644..148P} and VERITAS \citep{2009arXiv0911.0740V} 
with which we can directly compare our upper limits.

The WHIPPLE Collaboration observed the Perseus galaxy cluster \citep{2006ApJ...644..148P} 
for $\sim13$~hr obtaining an integral upper limit above $400$~GeV of 
$4.53\times10^{-12}$~cm$^{-2}$~s$^{-1}$ assuming a spectral index 
$\Gamma = -2.1$. We can compare this value with our integral upper
limit above $400$~GeV of $1.83\times10^{-12}$~cm$^{-2}$~s$^{-1}$
with $\Gamma = -2.2$ (see Table~\ref{tab:int_UL}). Our upper limit is 
significantly lower than the WHIPPLE one; clearly, this is
not a surprise as the MAGIC telescope belongs to a new generation
of IACTs. More recently, the VERITAS Collaboration observed Perseus 
\citep{2009arXiv0911.0740V} for $\sim8$~hr and obtained an integral 
upper limit above $126$~GeV of $1.27\times10^{-11}$~cm$^{-2}$~s$^{-1}$
assuming $\Gamma = -2.5$. We can compare this value with our corresponding
integral upper limit above $100$~GeV of $7.52\times10^{-12}$~cm$^{-2}$~s$^{-1}$ 
(see Table~\ref{tab:UL_resume}). Despite the fact that the VERITAS sensitivity of about 
$1$\% of Crab Nebula \citep{2009arXiv0907.4826N} is better than the MAGIC one, 
our upper limit is slightly lower than that found by \cite{2009arXiv0911.0740V} as 
expected from the significant difference in observation time.

\section{Cosmic-Ray-Induced Emission}
\label{subsec:CR} 
We use the upper limits on the integrated flux (Table~\ref{tab:int_UL}) to put
constraints on the CR-to-thermal pressure distribution and pursue three
different approaches. (1) We perform high-resolution hydrodynamical simulations
of cluster formation and evolution in a cosmological framework that include CR
physics to predict the $\gamma$-ray emission and to obtain limits on the
CR-to-thermal pressure. (2) Following \citet{2004A&A...413...17P}, we use a
simplified approach that assumes a constant CR-to-thermal energy density, a
power-law spectrum in momentum, and compare the resulting CR-to-thermal pressure
limits to those obtained by other IACT observations. (3) We use the observed
luminosity of the radio mini-halo to place a lower limit on the expected
$\gamma$-ray flux in the hadronic model of the radio mini-halo. This translates
into a minimum CR pressure that is crucial for disentangling the emission
mechanism in the radio and provides a clear prediction for the expected
$\gamma$-ray flux.

Before doing so, we detail our cosmological simulations that we base our main
analysis on.  To this end, we investigated the spatial and spectral properties of
$\gamma$-ray emission in these simulations and refer the reader to the theory
papers for further details \citep[][Pinzke \& Pfrommer, in
prep.]{2008MNRAS.385.1211P,2008MNRAS.385.1242P}.

\subsection{Cosmological Simulations}
\label{sec:simulations}
Simulations were performed using the ``concordance'' cosmological cold DM model
with a cosmological constant ($\Lambda$CDM) motivated by First Year cosmological
constraints of \emph{Wilkinson Microwave Anisotropy Probe} (WMAP). The cosmological parameters of our model are:
$\Omega_\rmn{m} = \Omega_\rmn{dm} + \Omega_\rmn{b} = 0.3$, $\Omega_\rmn{b} =
0.039$, $\Omega_\Lambda = 0.7$, $h = 0.7$, $n = 1$, and $\sigma_8 = 0.9$. Here,
$\Omega_\rmn{m}$ denotes the total matter density in units of the critical
density for geometrical closure today, $\rho_\rmn{crit} (z=0)= 3 H_0^2 / (8 \pi
G)$. $\Omega_\rmn{b}$ and $\Omega_\Lambda$ denote the densities of baryons and
the cosmological constant at the present day, respectively. The spectral index of the
primordial power-spectrum is denoted by $n$, and $\sigma_8$ is the rms
linear mass fluctuation within a sphere of radius $8\,h^{-1}$Mpc extrapolated to
$z=0$.

Our simulations were carried out with an updated and extended version of the
distributed-memory parallel TreeSPH code GADGET-2
\citep{2001NewA....6...79S,2005MNRAS.364.1105S}.  Gravitational forces were
computed using a combination of particle-mesh and tree algorithms.  Hydrodynamic
forces were computed with a variant of the smoothed particle hydrodynamics (SPH)
algorithm that conserves energy and entropy where appropriate, i.e.~outside of
shocked regions \citep{2002MNRAS.333..649S}.  We have performed high-resolution
hydrodynamic simulations of a sample of galaxy clusters that span over one and a
half decades in mass and show a variety of dynamical states ranging from relaxed
cool core clusters to violent merging clusters.  Our simulated clusters have
originally been selected from a low-resolution DM-only simulation
\citep{2001MNRAS.328..669Y}. Using the `zoomed initial conditions' technique
\citep{1993ApJ...412..455K}, the clusters have been re-simulated with higher
mass and force resolution.  In high-resolution regions, the DM particles had
masses of $m_\rmn{dm} = 1.61 \times 10^9\,h_{70}^{-1}\,\rmn{M}_\odot$ and SPH
particles $m_\rmn{gas} = 2.4\times 10^8\,h_{70}^{-1}\,\rmn{M}_\odot$ so each
individual cluster is resolved by $8 \times 10^4$-$4\times 10^6$ particles,
depending on its final mass. The SPH densities were computed from 48 neighbours,
allowing the SPH smoothing length to drop at most to half of the value of the
gravitational softening length of the gas particles. This choice of the SPH
smoothing length leads to our minimum gas resolution of approximately $1.1\times
10^{10}\,h_{70}^{-1}\,\rmn{M}_\odot$.  For the initial redshift we chose
$1+z_\rmn{init}=60$.  The gravitational force softening was of a spline form
\citep[e.g.,][]{1989ApJS...70..419H} with a Plummer-equivalent softening length
that is assumed to have a constant comoving scale down to $z = 5$, and a
constant value of $7\,h_{70}^{-1}$kpc in physical units at later epochs.

These simulations included radiative hydrodynamics, star formation, supernova
feedback and followed CR physics using a novel formulation that followed the
most important injection and loss processes self-consistently while accounting
for the CR pressure in the equations of motion \citep{2006MNRAS.367..113P,
  2007A&A...473...41E,2008A&A...481...33J}.  To obtain predictions of the
GeV-TeV $\gamma$-ray emission from clusters, we used an updated version of the
CR physics in our code. It is capable of following the spectral evolution of the
CR distribution function by tracking multiple CR populations in each gaseous
fluid element; {\em each} of these populations is described by an amplitude, a
low-momentum cut-off, and a characteristic power-law distribution in particle
momentum with a distinctive slope that is determined by the acceleration process
at formation shocks or supernova remnants (Pinzke \& Pfrommer 2010, in prep.).
Adiabatic CR transport processes such as compression and rarefaction, and a
number of physical source and sink terms which modify the CR pressure of each
particle are modeled. The most important sources considered are diffusive shock
acceleration at cosmological structure formation shocks and optional injection
by supernovae while the primary sinks are thermalization by Coulomb
interactions and catastrophic losses by hadronic interactions. We note that the
overall normalization of the CR distribution scales with the maximum
acceleration efficiency at structure formation shock waves. Following recent
observations at supernova remnants \citep{2009Sci...325..719H} as well as
theoretical studies \citep{2005ApJ...620...44K}, we adopt a realistic value of
this parameter and assume that 50\% of the dissipated energy at strong shocks is
injected into CRs while this efficiency rapidly decreases for weaker shocks
\citep{2007A&A...473...41E}.

We computed the $\gamma$-ray emission signal and found that it obeys a {\em
universal spectrum and spatial distribution} (Pinzke \& Pfrommer 2010, in
prep.). This is inherited from the universal concave spectrum of CRs in galaxy
clusters that is caused by the functional form and redshift dependence of the
Mach number distribution of structure formation shocks that are responsible for
the acceleration of CRs \citep{2006MNRAS.367..113P}. The CR distribution has a
spectral index of $\Gamma \simeq -2.5$ at GeV energies and experiences a
flattening towards higher energies resulting in $\Gamma \simeq -2.2$ at energies
above a few TeV. Hence, the resulting $\gamma$-ray spectrum from CR induced
pion-decay shows a characteristic spectral index of $\Gamma\simeq-2.2$ in the
energy regime ranging from 100 GeV to TeV.  The {\em spatial distribution} of the CR
number density is mainly governed by adiabatic transport processes
\citep{2007MNRAS...378..385P} and similarly attains an approximate universal
shape relative to that of the gas density. These findings allow us to reliably 
model the CR signal from nearby galaxy clusters using their true density profiles 
as obtained by X-ray measurements that we map onto our simulated density profiles. 

In addition to CR protons, we modeled relativistic electrons that have been
accelerated at cosmological structure formation shocks (primary CR electrons)
and those that have been produced in hadronic interactions of CRs with ambient
gas protons (secondary CR electrons). Both populations of CR electrons
contribute to the $\gamma$-ray emission through Compton up-scattering photons
from the cosmic microwave background as well as the cumulative star light from
galaxies. It turns out that the pion-decay emission of the cluster dominates
over the contribution from both IC components -- in particular, for
relaxed systems \citep{2008MNRAS.385.1242P}.

In our optimistic CR model ({\em radiative physics with galaxies}), we
calculated the cluster total $\gamma$-ray flux within a given solid angle. In
contrast, we cut the emission from individual galaxies and compact
galactic-sized objects in our more conservative model ({\em radiative physics
without galaxies}). In short, the ICM is a multiphase medium consisting of a
hot phase which attained its entropy through structure formation shock waves
dissipating gravitational energy associated with hierarchical clustering into
thermal energy. The dense, cold phase consists of the true interstellar medium
(ISM) within galaxies and at the cluster center as well as the ram-pressure-stripped 
ISM.  These cold dense gas clumps dissociate incompletely in the ICM
due to insufficient numerical resolution as well as so far incompletely
understood physical properties of the cluster plasma.  All of these phases
contribute to the $\gamma$-ray emission from a cluster. To assess the bias
associated with this issue, we performed our analysis with both limiting cases
bracketing the realistic case.

\begin{figure}[t!]
\centering
\includegraphics[width=1.0\linewidth]{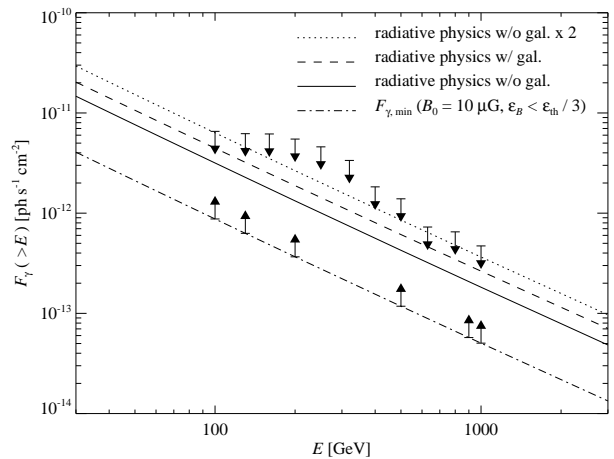}
\caption{Integral flux upper limits (this work, Table~\ref{tab:int_UL}) are
  compared with simulated integrated spectra of the $\gamma$-ray emission from
  decaying neutral pions that result from hadronic CR interactions with the
  ambient gas in the Perseus cluster. Our conservative model without galaxies
  (solid) is contrasted to our model with galaxies (dashed). We scaled our
  conservative model with a factor of 2 so that it is just consistent with
  the upper limits obtained in this work (dotted).  In our simulations, we
  assume an observationally motivated large value for the maximum CR energy
  injection efficiency at structure formation shocks and convert half of the
  dissipated energy to CRs at strong shocks. Smaller values would imply smaller
  $\gamma$-ray fluxes.  Additionally shown are minimum $\gamma$-ray flux
  estimates for the hadronic model of the radio mini-halo of the Perseus cluster
  (dash-dotted with minimum flux arrows; see the main text for details). Note that a
  non-detection of $\gamma$-rays at this level seriously challenges the hadronic
  model.}
\label{fig:ULs}
\end{figure}

In Figure~\ref{fig:ULs}, we compare the integral flux upper limits obtained in
this work (see Table~\ref{tab:int_UL}) with the simulated flux that is emitted
within a circle of radius $r_{0.6}=0^{\circ}.15$ for our two models, with and
without galaxies. The upper limits are a factor of 2 larger than our
conservative model and a factor of 1.5 larger than our most optimistic model
predictions implying consistency with our cosmological cluster simulations.  We
note however that our simulated flux represents a theoretical upper limit of the
expected $\gamma$-ray flux from structure formation CRs; lowering the maximum
acceleration efficiency would decrease the CR number density as well as the
resulting $\gamma$-ray emission.

\begin{figure*}[hbt!]
\centering
\includegraphics[width=0.42\linewidth]{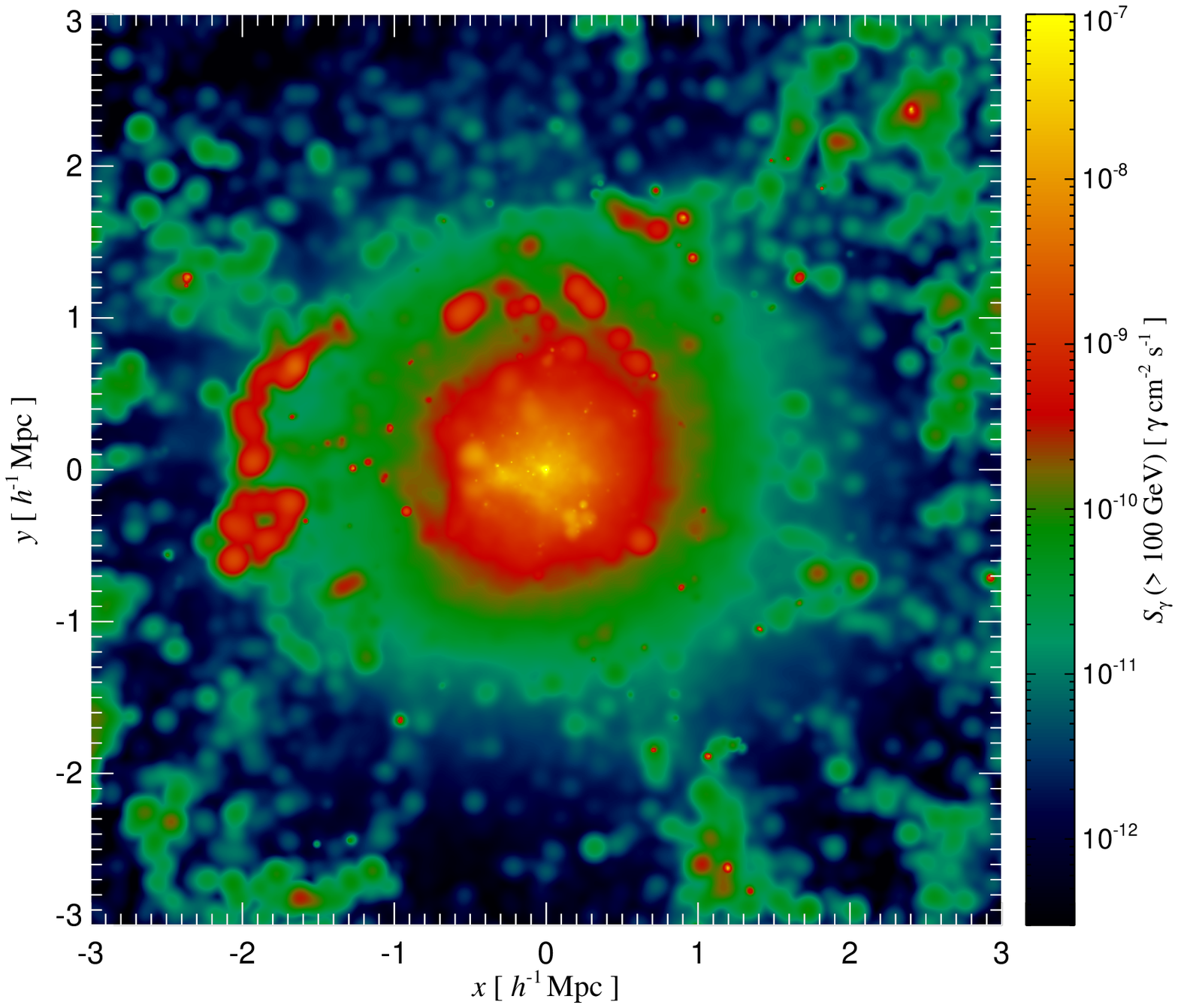}
\includegraphics[width=0.5\linewidth]{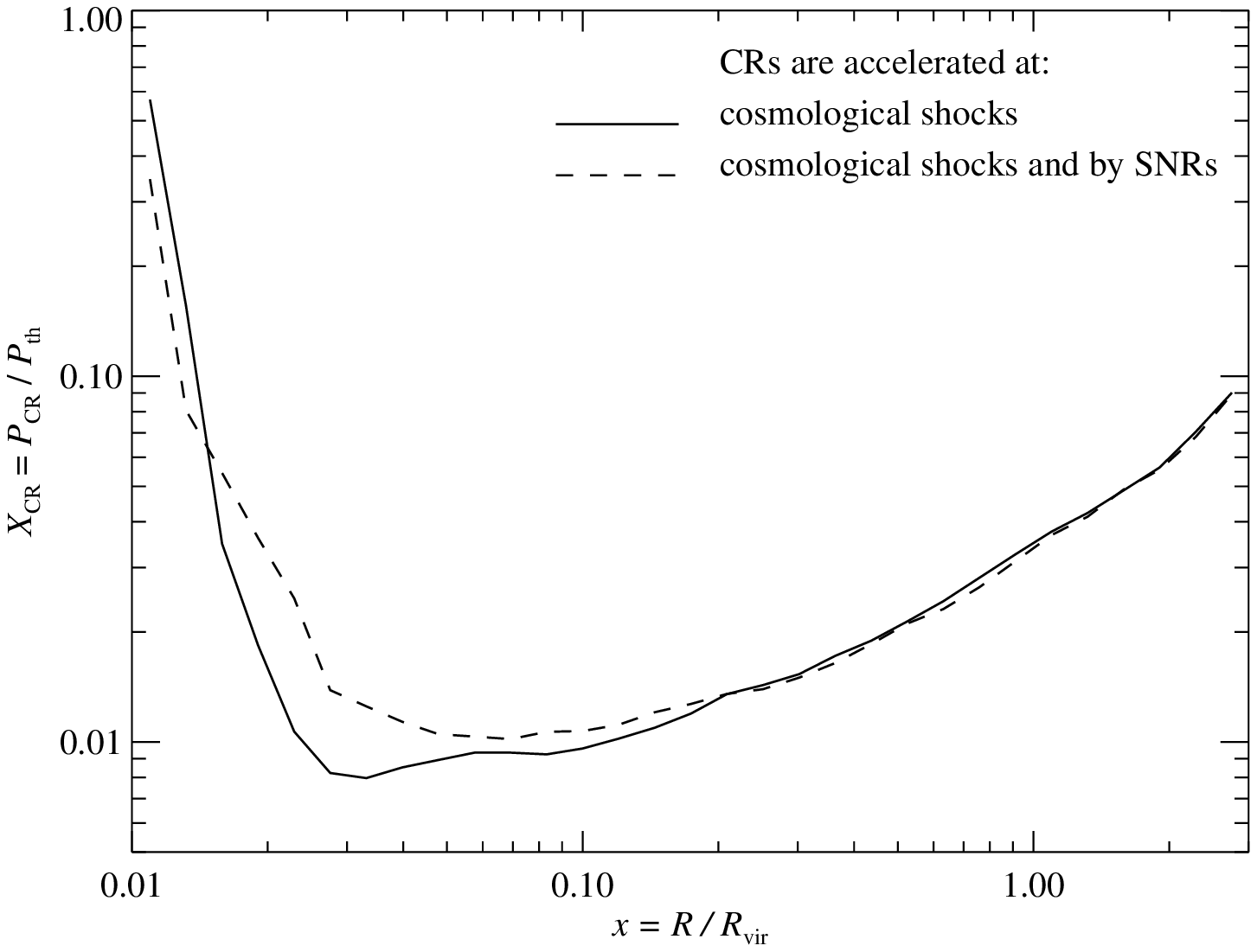}
\caption{Left: simulated $\gamma$-ray emission at energies $E>100$~GeV from a
  cluster that has twice the mass as Perseus (using the simulation of the
  cooling flow cluster g51 from \citealt{2008MNRAS.385.1211P}). We show the sum of
  pion-decay-induced $\gamma$-rays (which dominates the central and the total
  flux) and the IC emission of CR electrons accelerated at formation shocks and
  by hadronic CR interactions. Right: profile of the CR-to-thermal pressure
  (volume-weighted) of this cluster. We contrast a simulation where we only
  accelerate CRs at structure formation shocks of the entire cosmic history
  (solid) with one where we additionally account for CRs that are injected
  through supernova feedback within the star-forming regions in our simulation
  (dashed).}
\label{fig:XCR}
\end{figure*}

\subsection{Constraints on the Cosmic Ray Pressure}
\label{sec:CRp}
In Figure~\ref{fig:XCR}, we show the simulated $\gamma$-ray surface brightness
map of a cooling flow cluster of mass similar to Perseus. As the CR-induced
$\gamma$-ray flux is a radially declining function, so is the CR pressure. A
quantity that is of great theoretical interest is the CR pressure relative to
the thermal pressure $X_\CR = P_\CR / P_\rmn{th}$, as it directly assesses the
CR bias of hydrostatic cluster masses since the CR pressure enters in the
equation of motion.  On the right-hand side of Figure~\ref{fig:XCR}, we show the
profile of the CR-to-thermal pressure (volume-weighted) of this simulated
cluster. Moving from the periphery towards the center, this quantity is a
steadily declining function until we approach the cooling flow region around the
cD galaxy of this cluster (similar to NGC 1275) where the CR pressure rises
dramatically relative to that of the thermal gas which cools on a short time
scale \citep{2006MNRAS.367..113P}. The volume average is $\bra X_\CR\ket = \bra
P_\CR \ket / \bra P_\rmn{th}\ket = 0.02$, dominated by the region around the
virial radius, while the ratio of CR-to-thermal energy is given by
$E_\CR/E_\rmn{th} = 0.032$\footnote[2]{Note that for a CR population in clusters
  that have been accelerated in structure formation shocks, the relativistic
  limit $E_\CR/E_\rmn{th} = 2 \bra P_\CR \ket / \bra P_\rmn{th} \ket$ is not
  applicable since the CR pressure is dominated by the transrelativistic
  regime. This implies a somewhat harder equation of state for the CRs with a
  larger adiabatic index and yields the relation $E_\CR/E_\rmn{th} = 1.6 \bra
  P_\CR \ket / \bra P_\rmn{th} \ket$.}.  Perseus has a smaller mass and a
corresponding temperature that is only half of that of our simulated cooling
flow cluster. Noting that $X_\CR \propto 1/P_\rmn{th} \propto
1/kT$,\footnote[3]{This relation should only hold for regions with long thermal
  cooling times compared to the dynamical time scale. In particular, it breaks
  down toward the center of a cooling flow cluster where the thermal gas cools
  on a shorter time scale such that the forming cooling flow causes adiabatic
  contraction of the CR population. } we expect these values to be a factor of
$\simeq 2$ larger in Perseus, yielding $\bra X_\CR\ket \simeq 0.04$ for the
entire cluster and $\bra X_\CR\ket \simeq 0.02$ for the core region that we
probe with the present observation.

We have to scale our conservative model prediction by a factor of $\sim 2$ to
reach the upper limits (cf.~Figure~\ref{fig:ULs}) which implies that this work
constrains the relative pressure contained in CRs to $<8\%$ for the {\em entire
cluster} and to $<4\%$ for the {\em cluster core region}. The presence of dense 
gas clumps potentially biases the simulated $\gamma$-ray flux high and hence the 
inferred limits on $X_\CR$ low. Another source of bias could be unresolved 
point sources inside the cluster such as an active galactic nucleus (AGN). In 
the presented simulation of the 
cool core cluster g51, the bias due to subclumps amounts to a factor of 1.5 but it
could be as high as 2.4 which is the mean difference between our conservative
and optimistic model across our scaling relations.  We note however that the
latter case is already excluded by our upper limits provided the maximum shock
acceleration efficiency is indeed as high as 50\%. While there are indications
from supernova remnant observations of one rim region
\citep{2009Sci...325..719H} as well as theoretical studies
\citep{2005ApJ...620...44K} that support such high efficiencies, to date it is
not clear whether these efficiencies apply in an average sense to strong
collisionless shocks or whether they are realized for structure formation shocks
at higher redshifts.  Improving the sensitivity of the presented type of
observations will help in answering these profound plasma astrophysics
questions.

In Figure~\ref{fig:XCR}, we additionally compare a simulation where we only
accelerate CRs at structure formation shocks with one where we additionally
account for CRs that are injected through supernova feedback within the star-forming 
regions in our simulation. Outside the cD galaxy, there is no
significant difference visible which suggests that the CRs injected into the ICM
by supernova-driven winds are negligible compared with those accelerated by
structure formation shocks.  While this is partly an artifact of our simulations
that neglect CR diffusion, we expect this behavior due to the adiabatic losses
that CRs suffer as they expand from their compact galactic ISM into the dilute
ICM. Assuming a conservative value for the density contrast of $\Delta=10^{-3}$,
the CR pressure is diluted by $P_\CR \sim \Delta^{4/3}\,P_{\CR,\rmn{ISM}} \sim
10^{-4}\,P_{\CR,\rmn{ISM}}$.

\subsection{Simplified Approach and Comparison to Previous Results}
\label{subsec:comparison}

As anticipated in section~\ref{subsec:UL_diss}, there are few existing IACT
observations of galaxy clusters; some of which derived limits on the
CR-to-thermal pressure contained in clusters, in particular the WHIPPLE
observation of the Perseus cluster \citep{2006ApJ...644..148P} and the HESS
observations of the Abell~85 \citep{2009A&A...495...27A,2009arXiv0907.3001D} and
Coma \citep{2009arXiv0907.0727T} clusters. These works used simplifying
assumptions about the spectral and spatial distribution of CRs. They typically
assumed a single CR power-law distribution with a spectral index of
$\Gamma=-2.1$ (that provides optimistic limits on the CR-to-thermal pressure)
and assumed that the CR energy density is a constant fraction of the thermal
energy density throughout the entire cluster. Based on these two assumptions,
WHIPPLE and HESS found in Perseus and Abell~85 $E_\CR/E_\rmn{th} <0.08$,
respectively, while HESS found $E_\CR/E_\rmn{th}< 0.2$ in Coma.

To facilitate comparison with these earlier works, we repeated the data analysis
with a spectral index $\Gamma = -2.1$ to obtain an integral upper limit
$\mathcal{F}_{\rmn{UL}} (>100 \mbox{ GeV}) = 6.22\times 10^{-12} \mbox{
  cm}^{-2}\mbox{ s}^{-1}$. Following the formalism of
\citet{2004A&A...413...17P}, we compute the $\gamma$-ray flux of a CR population
with $\Gamma = -2.1$ within a circular region of radius $r_{0.6}=0^{\circ}.15$ or
equivalently 200~kpc. In our {\em isobaric model of CRs} we assume that the CR
pressure scales exactly as the thermal pressure and constrain $E_\CR /
E_\rmn{th} < 0.053$ which corresponds to an averaged relative pressure of $\bra
X_\CR\ket = \bra P_\CR \ket / \bra P_\rmn{th}\ket = 0.033$. This would be the
most stringent upper limit on the CR energy in a galaxy cluster.

In our {\em adiabatic model of CRs} we account for the centrally enhanced CR
number density due to adiabatic contraction during the formation of the cooling
flow \citep{2004A&A...413...17P}.  We assume that the CRp population scaled
originally as the thermal population but was compressed adiabatically during the
formation of the cooling flow without relaxing afterward (we adopted
temperature and density profiles given by \citealp{2003ApJ...590..225C}).  In this
model, we obtain an enhanced $\gamma$-ray flux level for virtually the same
volume-averaged CR pressure or vice versa for a given flux limit; hence, we can
put a tighter constraint on the averaged CR pressure. We constrain $E_\CR /
E_\rmn{th} < 0.03$ which corresponds to an averaged relative pressure of $\bra
X_\CR\ket = \bra P_\CR \ket / \bra P_\rmn{th}\ket = 0.019$.

How can we reconcile these tighter limits with our simulation-based slightly
weaker limit? We have to compare our simulated CR profile to a CR distribution
that does not show any enhancement relative to the gas density. In the central
region for $r<200$~kpc, we derive an adiabatic compression factor of 1.7 that
matches that in our simplified approach -- suggesting that our simple adiabatic
model captures the underlying physics quite realistically. Second, we have
then to relate the pressure of a power-law spectrum with $\Gamma=2.1$ to our
simulated concave spectrum.  Noting that the $\gamma$-rays at 100~GeV are
produced by CR protons at $\simeq 1$~TeV, we normalize both spectra at 1~TeV and
find that the simulated spectrum contains a larger pressure by a factor of
1.8. This factor brings the limit of our simplified adiabatic model into
agreement with our simulation-based limit of the relative CR pressure $\bra
X_\CR\ket< 4\%$ for the {\em cluster core region}. Finally, since $\gamma$-ray
observations are only sensitive to the cluster core regions (the emission
is expected to peak in the center due to the high target gas densities), they
cannot constrain the average CR-to-thermal pressure within the entire
cluster. Hence we have to use cosmological cluster simulations to address how
much CR-to-thermal pressure could be additionally hidden in the peripheral
cluster regions.

\subsection{Minimum $\gamma$-ray Flux}
For clusters that host radio (mini-)halos we are able to derive a minimum
$\gamma$-ray flux in the hadronic model of CR interactions. The idea is based on
the fact that a steady-state distribution of CR electrons loses all its energy
to synchrotron radiation for strong magnetic fields ($B \gg B_\rmn{CMB} \simeq
3.2 \mu$G) so that the ratio of $\gamma$-ray to synchrotron flux becomes
independent of the spatial distribution of CRs and thermal gas
\citep{2008MNRAS.385.1242P}.  This can be easily seen by considering the pion-decay-induced 
$\gamma$-ray luminosity $L_\gamma$ and the synchrotron luminosity
$L_\nu$ of a steady-state distribution of CR electrons that has been generated
by hadronic CR interactions:
\begin{eqnarray}
L_\gamma &=& A_\gamma \dps\int\dd V\, n_\CR n_\rmn{gas},\\
   L_\nu &=& A_\nu \dps\int \dd V\, n_\CR n_\rmn{gas} \frac{\dps
    \eps_B^{(\alpha_\nu+1)/2}}{\dps\eps_\rmn{CMB}+\eps_B}\\
  &\simeq&A_\nu \dps\int \dd V\, n_\CR n_\rmn{gas}\quad\rmn{for}~\eps_B \gg \eps_\rmn{CMB}.
\end{eqnarray}
Here $A_\gamma$ and $A_\nu$ are dimensional constants that depend on the
hadronic physics of the interaction \citep{2008MNRAS.385.1211P,
  2008MNRAS.385.1242P} and $\alpha_\nu\simeq 1$ is the observed synchrotron
spectral index. Hence, we can derive a minimum $\gamma$-ray flux in the hadronic
model:
\begin{equation}
  \label{eq:Fmin}
  \mathcal{F}_{\gamma,\rmn{min}} = 
  \frac{\dps A_\gamma}{\dps A_\nu}\frac{\dps L_\nu}{\dps 4\pi D_\rmn{lum}^2},
\end{equation}
where $L_\nu$ is the observed luminosity of the radio mini-halo and
$D_\rmn{lum}$ denotes the luminosity distance to the respective cluster.
Lowering the magnetic field would require an increase in the energy density of
CR electrons to reproduce the observed synchrotron luminosity and thus increase
the associated $\gamma$-ray flux.

Using the values of Table~\ref{tab:data}, we obtain a minimum $\gamma$-ray flux
in the hadronic model of the radio mini-halo of $\mathcal{F}_{\gamma, \rmn{min}}
(>100 \mbox{ GeV}) = 6\times 10^{-13} \mbox{cm}^{-2}\mbox{ s}^{-1}$, assuming a
power-law CR distribution with $\Gamma \gsim -2.3$. This lower limit is
independent of the spatial distribution of CRs and magnetic fields. We note that
the spectral index is consistent with the radio data\footnote[4]{The CR protons
  responsible for the GHz radio emitting electrons are $\sim$ 100 times less
  energetic than those CR protons that are responsible for the TeV $\gamma$-ray
  emission. This is consistent with the concave curvature found in the CR
  spectrum by Pinzke \& Pfrommer (2010, in prep).}. It turns out that the requirement
of strong magnetic fields violates the energy conditions in clusters as it
implies a magnetic energy density that is larger than the thermal energy density
-- in particular, at the peripheral cluster regions. The minimum $\gamma$-ray
flux condition requires a constant (large) magnetic field strength throughout
the cluster while the thermal energy density is decreasing by more than a factor
of 100 from its central value. This would imply that the magnetic field
eventually dominates the energy density at the virial regions -- a behavior that
is unstable as it is subject to Parker-like buoyancy
instabilities. Additionally, such a configuration would be impossible to achieve
in first place as the magnetic energy density typically saturates at a fixed
fraction of the turbulent energy density which itself is only a small fraction
of the thermal energy density in clusters \citep{2004A&A...426..387S}.  Hence,
these considerations call for lowering the assumed cluster magnetic fields which
should strengthen the lower limits on the $\gamma$-ray flux considerably --
however at the expense that these limits inherit a weak dependence on the
spatial distribution of magnetic fields and CRs.

Estimates of magnetic fields from Faraday rotation measures (RMs) have undergone
a revision in the last few years with more recent estimates typically in the
order of a few $\mu$G with slightly higher values up to 10 $\mu$G in cooling
flow clusters \citep{2004JKAS...37..337C,2006A&A...453..447E}. For the Perseus
radio mini-halo, Faraday RMs are available only on very small scales
\citep{2006MNRAS.368.1500T}, i.e.~few tens of pc.  RM estimates are of the order
of $\sim7000 \mbox{ rad m}^{2}$ leading to magnetic field values of $\sim25
\mu$G assuming that the Faraday screen is localized in the ICM. This, however,
appears to be unlikely as variations of $10\%$ in the RM are observed on
pc-scales \citep{2002MNRAS.334..769T}, while ICM magnetic fields are expected to
be ordered on significantly larger scales of a few kpc
\citep{2006MNRAS.368.1500T,2005A&A...434...67V,
  2006A&A...453..447E}. Application of the classical minimum-energy argument to
the Perseus radio mini-halo data leads to estimates for the central magnetic
field strength of $B_0\simeq 7 \mu$G or even $B_0\simeq 9 \mu$G for the more
appropriate hadronic minimum-energy argument \citep{2004MNRAS.352...76P}.

We select a cooling flow cluster of our sample that is morphologically similar
to Perseus with a mass $M_{200} \simeq 10^{15}\,\rmn{M}_\odot$ (the simulated
cluster g51 of \citealp{2008MNRAS.385.1211P}).  We adopt a conservative choice
for the central magnetic field strength of $\sim10 \mu$G and parameterize the
magnetic energy density in terms of the thermal energy density by $\eps_B
\propto \eps_\rmn{th}^{0.5}$ which ensures $\eps_B < \eps_\rmn{th}/3$ in the
entire cluster. This allows us to strengthen the physically motivated lower
limit to $\mathcal{F}_{\gamma, \rmn{phys.\,min}} (> 100 \mbox{ GeV}) = 8.5\times
10^{-13} \mbox{ cm}^{-2} \mbox{ s}^{-1}$ as shown by the dash-dotted line in
Figure~\ref{fig:ULs}.  In the hadronic model, this minimum $\gamma$-ray flux
implies a minimum CR pressure relative to the thermal pressure.
Figure~\ref{fig:ULs} shows that the minimum flux $\mathcal{F}_{\gamma,
  \rmn{phys.\,min}}$ is a factor of 3.6 lower than the simulated flux for
Perseus in our conservative model. As seen in Sect.~\ref{sec:CRp}, this model
corresponds to a relative CR pressure of $\bra X_{\CR}\ket = \bra P_{\CR}\ket /
\bra P_\rmn{th}\ket = 0.04$ where the averages represent volume averages across
the entire cluster. Hence we obtain a minimum relative CR pressure, $\bra
X_{\CR, \,\rmn{min}}\ket = \bra P_{\CR, \,\rmn{min}}\ket / \bra
P_\rmn{th}\ket/3.6 = 0.01$. This minimum CR pressure corresponds to a minimum
total CR energy of $E_{\CR\,\rmn{min}} = E_{\CR\,\rmn{min}} / E_\rmn{th}\times
E_\rmn{th} = 1.6\, \bra X_{\CR, \,\rmn{min}} \ket \times E_\rmn{th} = 9\times
10^{61}\,\rmn{erg}$ where we integrated the temperature and density profiles
from X-ray observations \citep{2003ApJ...590..225C} to obtain the total thermal
energy of $E_\rmn{th} =5.7\times 10^{63}\,\rmn{erg} $. These considerations show
the huge potential of combining future TeV $\gamma$-ray and radio observations
in constraining physical models of the non-thermal cluster emission and of
obtaining important insights into the average distribution of cluster magnetic
fields.

\section{Dark Matter Annihilation}
\label{subsec:DM}  
As discussed in Section~\ref{section:target_DM}, the expected DM annihilation
flux is proportional to the product of a factor that encloses all the particle
physics and a second one that accounts for all the involved
astrophysics. Therefore, in order to obtain an estimate of the annihilation
flux, we need to choose a particular particle physics model (that was not needed
in Section~\ref{section:target_DM}, since only a comparative study was done
there) in addition to the modeling of the DM distribution.  Although the
uncertainties in the particle physics factor $f_{\mathrm{SUSY}}$ are very large
and spread over some orders of magnitude \citep[see, 
e.g.,][]{2008ApJ...679..428A}, it is common to use the most optimistic value for
a given energy threshold of the telescope. This factor just acts as a rescaling
factor in the total flux, so we could change to the other particle physics model
simply by rescaling for its new value.  Let us assume $f_{\mathrm{SUSY}} =
10^{-32}$~GeV$^{-2}$~cm$^{3}$~s$^{-1}$ above 100~GeV, which corresponds to one
of the most optimistic allowed scenarios at the energies of interest here
\citep{2007PhRvD..76l3509S}, with the neutralino as a DM particle. Then, taking a
value of $1.4\times10^{16}$~GeV$^2$~cm$^{-5}$ for the integrated astrophysical
factor inside $r_s$ (as given in Section~\ref{section:target_DM}), we obtain a
maximum DM annihilation flux of $\sim1.4\times10^{-16}$~cm$^{-2}$~s$^{-1}$ for
energies above 100~GeV.  The comparison with the derived upper limits from our
observations is not very constraining. Assuming a generic DM annihilation
spectrum without a cutoff and a spectral index -1.5 as a good approximation
\citep[e.g.,][]{2008ApJ...679..428A,2009ApJ...697.1299A}, it can be seen from
Table~\ref{tab:UL_resume} that we need a boost in flux in the order of 10$^4$ to
reach the predicted DM annihilation flux values, since $\mathcal{F}_{\rmn{UL}}$ ($>$100 GeV) =
$4.63\times10^{-12}$~cm$^{-2}$~s$^{-1}$.

This boost factor could come from different uncertainties that may enhance the annihilation $\gamma$-ray flux 
notably and that were not taken into account in the above calculation. One of them, the presence of substructures, 
could play a crucial role for Perseus, as explained in section~\ref{section:target_DM}. Although still uncertain, 
its effect could enhance the expected annihilation flux by more than a factor of $10$ for Perseus-size halos 
according to \cite{2008ApJ...686..262K}. More recent work has shown that the expected boost factors could be 
as high as $200$ \citep{2008MNRAS.391.1685S,2008Natur.456...73S}. However, with IACTs it is challenging 
to make use of these large boost factors as their contribution is expected to be more important on large angular 
scales comparable to the virial extent of the cluster. Detailed modeling of the substructures is needed in 
order to correctly evaluate their impact on the Perseus DM-induced signal.  
Finally, also recently proposed mechanisms in the particle physics side, such as the internal bremsstrahlung 
\citep{2008JHEP...01..049B} and the Sommerfeld effect \citep{2009PhRvD..79h3523L,2009arXiv0905.1948P}, 
could enhance the DM annihilation flux by more than one order of magnitude for some particle physics 
models.

It is worth noting that the result obtained here for the boost factor needed in order to probe the
predicted DM annihilation flux is comparable with previous observations of the Milky Way satellite galaxies
\citep{2008ApJ...679..428A,2009ApJ...697.1299A}.

\section{The NGC~1275 Emission}
\label{subsec:AGN}
\begin{figure}[hbt!]
\centering
\includegraphics[width=1.05\linewidth]{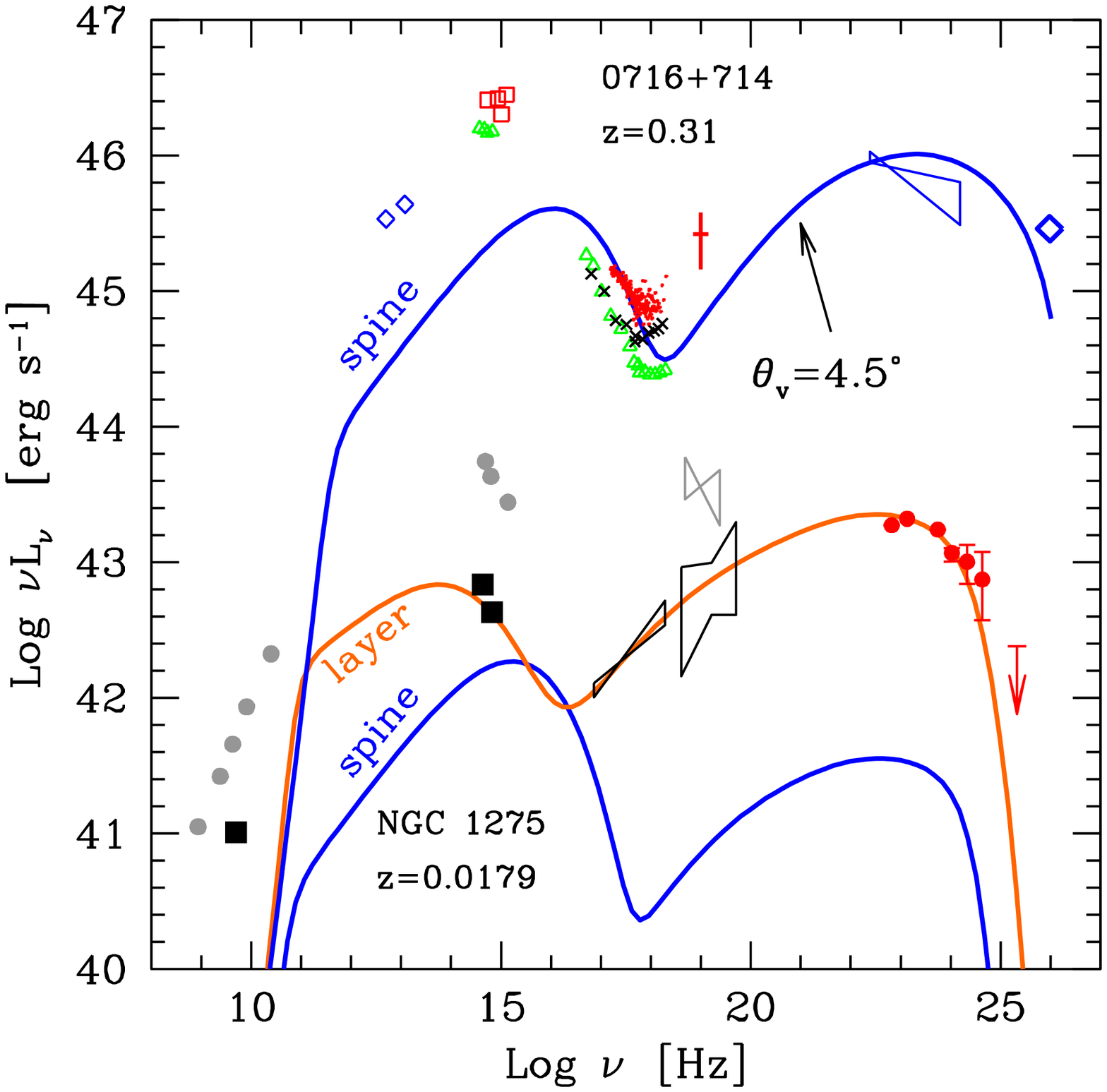}
\caption{SED of the NGC~1275 core (lower two lines and data) and that of the
  well known blazar S5~0716+714 for comparison (upper line and data). Gray
  filled circles are data points in the radio and optical bands from
  \cite{2009arXiv0904.1904T}. Filled black squares show, instead, the radio
  (VLBI; \citealp{1996ApJS..107...37T}) and the optical emission (Hubble Space Telescope (HST);
  \citealp{1999A&A...349...77C}) of the core alone. The soft X-ray bow tie is
  from \emph{Chandra} \citep{2006A&A...451...35B}, while the red filled circles
  represent the \emph{Fermi}-LAT spectrum taken from \cite{2009arXiv0904.1904T}. The red
  arrow shows the MAGIC upper limit between 80 and 100 GeV. The lower blue and
  red lines show the emission from the spine and the layer of the structured
  jet, respectively. The upper blue lines is the SED of the spine as observed at a small angle
  (see the text for details); for comparison, we report historical data of
  S5~0716+714 (data from \citealp{2009MNRAS.394L.131T} and references therein).}
\label{fig:SED_NGC}
\end{figure}

The SED of the NGC~1275 core is shown in Figure~\ref{fig:SED_NGC}. The radio 
and optical data represented with gray filled circles \citep{2009arXiv0904.1904T} 
have been obtained with low resolution and thus include a large contribution from 
the large-scale regions of the jet (radio) and from the host galaxy (optical). 
In the following, we model the data corresponding to the core emission. 
This is different to what was done by \cite{2009arXiv0904.1904T} who used the 
low resolution data in their models. We calculated our upper limit, shown in Figure~\ref{fig:SED_NGC} 
as a red arrow, assuming that the spectrum in the MAGIC energy band is a power law with spectral 
index $\Gamma=-2.5$, as indicated by the extrapolation of the last points of the \emph{Fermi}-LAT spectrum. Note, 
however, that the level of the differential upper limits is only weakly dependent on the assumed spectral 
index (see Table~\ref{tab:dif_UL}).

The data clearly show a double-peak SED. The radio-optical data suggest 
a peak of the emission in the IR band, similar to the case of other 
$\gamma$-ray emitting radio galaxies \citep{2008MNRAS.385L..98T,2009MNRAS.394L.131T}. 
High-energy data constrain the peak frequency of the second component at 
about 100 MeV. As discussed in \cite{2008MNRAS.385L..98T}, in these 
cases a one-zone synchrotron self-Compton (SSC) model for the entire emission 
is implausible, since the large separation in frequency between the two peaks would 
require extremely large values of the Doppler factor:
\begin{equation}
\delta\simeq 258 \, L_{\rm s,42.8} L_{\rm C,43.4}^{-1/2} \, \nu_{\rm s,13.5}^{-2} \, 
\nu_{\rm C,23} \, R_{16}^{-1}
\end{equation}
where  $L_{\rm s} = \nu_{\rm s} L(\nu_{\rm s})$, $L_{\rm C} = \nu_{\rm c} L(\nu_{\rm c})$, 
$\nu_{\rm s}$ and $\nu_{\mathrm C}$ are the synchrotron and SSC peak luminosities 
and frequencies, respectively, and $R$ is the size of the emitting region.
Here $Q=10^x Q_x$ in cgs units, and we use the values derived for NGC~1275. 
In this estimate, we assume the typical size of the emission regions derived in 
blazars, $R=10^{16}$ cm, though the \emph{Fermi}-LAT data do not allow us to constrain 
the radius of the emission region using the variability \citep{2009arXiv0904.1904T}.
Such large values of the Doppler factor are rather unlikely. Typical values found modeling 
the SED of blazars are around $10-20$ (e.g., \citealp{2008MNRAS.385..283C}), with few 
extreme TeV BL Lacs requiring larger values during exceptional states ($\delta\sim50-100$; 
e.g., \citealp{2008MNRAS.384L..19B,2008MNRAS.386L..28G}). Arguments based on the observation 
of  superluminal motions at Very Long Baseline Interferometry (VLBI) scales (e.g., \citealp{2004ApJ...609..539K}) and on the unification 
of blazars with radio galaxies also require values around 10 (e.g., \citealp{1995PASP..107..803U}).

The most direct way to overcome the problem posed by the large Doppler 
factor is to assume two emission regions, as in the spine-layer model 
of \cite{2005A&A...432..401G}. In this scenario the jet is assumed to be 
structured, with a fast inner region (the spine) surrounded by a slower 
sheet (the layer). Both components produce synchrotron and IC 
radiation and they are in radiative interplay: the synchrotron 
radiation from one component is seen boosted (by the relative velocity) 
by the other one and thus the IC emission of both regions is enhanced 
with respect to the standard SSC. In radio galaxies, in which the jet 
is observed at relatively large angles, the emission is expected to be 
dominated by the layer, which, due to the lower bulk Lorentz factor,  
has a larger emission cone. At a smaller angle, instead, the emission is 
dominated by the spine, as in blazars.

We reproduce the SED with the spine-layer model. The orange line in Figure~\ref{fig:SED_NGC} 
shows the emission from the layer, while the spine produces the emission shown 
by the blue bottom line. The spine is assumed to be a cylinder of radius 
$R=1.5\times 10^{16}$~cm, height $H_{\rm S}=1.5\times 10^{16}$~cm 
(as measured in the spine frame), and in motion with bulk Lorentz 
factor $\Gamma_{\rm S}=15$. The layer is modeled as an hollow 
cylinder with internal radius $R$, external radius $R_2=1.2\times R$, 
height $H_{\rm L}=4\times 10^{16}$~cm (as measured in the frame of 
the layer), and bulk Lorentz factor $\Gamma_{\rm L}=3$. Each region 
contains a tangled magnetic field with intensity $B_{\rm S}=2.5$~G and 
$B_{\rm L}=1$~G, and it is filled by relativistic electrons assumed to 
follow a (purely phenomenological) smoothed broken power-law 
distribution extending from $\gamma_{\rm min}$ to $\gamma_{\rm max}$ 
and with indices $n_1$, $n_2$ below and above the break at $\gamma_{\rm b}$.  
For the spine we use  $\gamma_{\rm min}=40$, $\gamma _{\rm b}=2\times 10^4$,  
$\gamma_{\rm max}=10^5$, $n_1=2$, $n_2=3.5$. For the layer $\gamma_{\rm min}=10$, 
$\gamma _{\rm b}=4\times 10^3$,  $\gamma_{\rm max}=10^5$, $n_1=2.4$, $n_2=4.2$. 
The normalization of these distributions is calculated assuming that the 
systems produce an assumed (bolometric) synchrotron luminosity 
$L^\prime_{\rm syn, S}=10^{42}$~erg~s$^{-1}$  and $L^\prime_{\rm syn,L}=2.7 \times 10^{41}$~erg~s$^{-1}$ 
(as measured in the local comoving frame of the spine and layer, respectively),
which is an input parameter of the model.
As said above, the seed photons for the IC scattering are 
not only those produced locally in the spine (layer), but we also consider 
the photons produced in the layer (spine). We assume a viewing angle of $\theta =15^{\circ}$.
As discussed above, the same jet observed at a smaller angle would be dominated 
by the emission from the spine and we expect that its SED resembles those of 
typical blazars. We show the SED of the jet when observed at an angle of $4^{\circ}.5$ 
(blue upper line in Figure~\ref{fig:SED_NGC}). The SED is dominated 
by the emission of the spine. For comparison, we report historical data for 
the well-known blazar S5~0716+714 (data from \citealp{2009MNRAS.394L.131T} 
and references therein). 

Note that, as observed, the model naturally predicts a very rapid decrease of the
emission level above 10~GeV, due to the decreasing efficiency of the IC scattering in the 
Klein-Nishina regime. The position of this break is tightly constrained by the \emph{Fermi}-LAT 
spectrum and MAGIC upper limit. In our model, this is critically dependent on the value of the 
frequency of the target photons for the IC scattering that in the spine-layer scenario are mainly 
those coming from the spine (and scattered by the electrons in the layer). Therefore, the determination
of the cut-off frequency between the \emph{Fermi}-LAT and the MAGIC band allows us to infer the peak 
frequency of the synchrotron component of the spine. For instance, assuming that the peak of the 
spine is at IR frequencies or below (using for the layer the same parameters adopted above), we predict a 
flux in the MAGIC band above the measured upper limit. This argument allows us to fix the synchrotron peak 
of the spine at optical-UV frequencies. This, in turn, assures that the beamed counterpart of NGC~1275 is an 
intermediate BL~Lac object, as the chosen S5~0716+714. In conclusion, the knowledge of the upper limit at 
the low-energy end of the MAGIC band offers us the important possibility of having independent limits on 
the characteristics of the emission of the (otherwise invisible) spine and thus of constraining the kind of beamed 
counterpart of this radio galaxy. Future observations can confirm or rule out our interpretation. In particular, 
the detection of photons above $\sim 100$~GeV would be challenging for the scenario depicted here, requiring 
major changes in the emission properties of the spine.

\section{Conclusions}
\label{sec:conclusions}
The Perseus cluster was observed by MAGIC during 2008 November and December 
resulting in 24.4~hr effective observation time of very high data 
quality. No significant excess of $\gamma$-ray was detected beyond the energy 
threshold of 80~GeV. 

Using simplified assumptions (power-law CR spectra, constant ratio of
CR-to-thermal energy density) that have been adopted in earlier work, we obtain
a limit on the CR energy of $E_\CR / E_\rmn{th} < 5\%$. This limit could be
tightened furthermore by considering an adiabatically contracted CR population
during the formation of the cooling flow yielding $E_\CR / E_\rmn{th} < 3\%$.
This would be the most stringent constraint on the CR energy using $\gamma$-ray
observations to date. Using cosmological cluster simulations, it turns out that
these assumptions are not fulfilled for CR populations that have been
accelerated by structure formation shocks: while the adiabatic model seems to
match the simulated CR profiles toward the center very well, the expected ratio
of CR-to-thermal pressure increases toward the peripheral cluster regions
causing the volume-averaged pressure across the entire cluster to increase by a
factor of 2. In addition, the CR spectral distribution shows a concave curvature
with a spectrum that flattens toward high energies with a spectral index of
$\Gamma\simeq-2.2$ in the TeV regime.  This implies that the CR pressure is
enhanced by an additional factor of almost 2. Using our simulated flux, we
obtained an upper limit on the CR-to-thermal pressure averaged across the {\em
  entire cluster volume} of $\bra X_\CR \ket < 8\%$ and $<4\%$ for the {\em
  cluster core region}.  This corresponds to an upper limit on the CR energy of
$E_\CR / E_\rmn{th} < 13\%$ and $<6.5\%$, respectively. We note that this is the
first work where results from cosmological simulations and observational data
analysis are combined.  This demonstrates the need for cosmological simulations
in order to more reliably predict CR spectra which provide a safeguard against
too simplified and optimistic models which then lead to limits that are too
tight.

The upper limits resulting from the data analysis are a factor of $\simeq 2$
larger than our conservative model prediction for the CR-induced $\gamma$-ray
emission and hence in agreement with our cosmological cluster simulations.
Future more sensitive measurements will be able to put interesting constraints
on the maximum shock acceleration efficiency.  Using minimum $\gamma$-ray flux
arguments, we show that improving the sensitivity of this observation by a
factor of about 7 will enable us to finally critically test the hadronic model for
the Perseus radio mini-halo: a non-detection of $\gamma$-ray emission at this
level implies CR fluxes that are too small to produce enough electrons
through hadronic interactions with the ambient gas protons to explain the
observed synchrotron emission.

As DM dominates the cluster mass, significant $\gamma$-ray emission resulting
from its annihilation is also expected. With the assumed particle physics model,
one of the most optimistic allowed scenarios \citep{2007PhRvD..76l3509S} with
the neutralino as a DM particle, the boost factor for the typically expected DM
annihilation-induced emission is constrained to $<10^4$. Note that for this 
estimation, we neglected possible contributions from internal bremsstrahlung, 
Sommerfeld enhancement as well as enhancement factors due to substructures.

The upper limits obtained for the NGC~1275 emission are consistent with the
recent detection by the \emph{Fermi}-LAT satellite. In this case a one-zone SSC model
for the entire emission is implausible, since the large separation in frequency between
the two peaks would require extremely large values of the Doppler factor for the jet
\citep{2008MNRAS.385L..98T}. The most direct way to overcome this problem is to assume 
two emission regions, as in the spine-layer model \citep{2005A&A...432..401G} which explains 
the radio galaxy emission.

While no galaxy cluster has been detected in $\gamma$-rays up to now, our
estimations indicate that Perseus is among the most promising clusters to be
detected by IACTs. Using the newly inaugurated MAGIC second telescope and
operating the telescopes in stereo mode \citep{2009arXiv0907.0960C}, a total observation 
time of about 150~hr may give us a chance to detect the CR-induced $\gamma$-ray emission or
to definitively probe the validity of the hadronic model of radio
(mini-)halos. As the emission of NGC~1275 dominates the accessible energy range
of the \emph{Fermi}-LAT satellite, it could potentially hinder the satellite from
detecting the CR as well as the DM-induced $\gamma$-ray emission in this
cluster. Similar problems might arise in other clusters. Therefore, the IACTs
will play a crucial role in the quest for $\gamma$-ray emission from galaxy
clusters.

\acknowledgments
We thank the anonymous referee for valuable comments.
We thank Volker Springel for the important work on CR 
implementation into the GADGET code and comments on the paper. 
We also thank Yoel Rephaeli for the useful comments on the paper and 
Eugene Churazov for providing us \emph{XMM-Newton} X-ray contours. 
We thank the Instituto de Astrofisica de Canarias for the excellent 
working conditions at the Observatorio del Roque de los Muchachos in 
La Palma. The support of the German BMBF and MPG, the Italian INFN,
and Spanish MICINN is gratefully acknowledged. 
This work was also supported by ETH Research Grant 
TH 34/043, by the Polish MNiSzW Grant N N203 390834, 
by the YIP of the Helmholtz Gemeinschaft, and by grant DO02-353
of the Bulgarian National Science Fund.\\

\bibliographystyle{apj}
\bibliography{paper}

\end{document}